\documentstyle[12pt]{article}
\begin{document}
\setlength{\unitlength}{1mm}

{\hfill Preprint DSF-8/94}\\

{\hfill April 1994}\\
\begin{center}
{\Large\bf Semiclassical Gravitational Effects in de Sitter Space}\\
\end{center}
\begin{center}
{\Large\bf at Finite Temperature}
\end{center}

\bigskip\bigskip

\begin{center}
{{\bf D.V. Fursaev$^{1,2}$} and {\bf G. Miele$^{2}$}}
\end{center}

\begin{center}
{$^{1}$ {\it Laboratory of Theoretical Physics, Joint Institute for
Nuclear Research, \\
Head Post Office, P.O.Box 79, Moscow, Russia}}
\end{center}

\begin{center}
{$^{2}$ {\it Dipartimento di Scienze Fisiche, Universit\`a di Napoli -
Federico II -, and INFN\\
Sezione di Napoli, Mostra D'Oltremare Pad. 19, 80125, Napoli, Italy}}
\end{center}

\bigskip\bigskip\bigskip

\begin{abstract}
In the framework of finite temperature conformal scalar field theory
on de Sitter space-time the linearized Einstein equations for the
renormalized stress tensor are exactly solved. In this theory
quantum field fluctuations are concentrated near two spheres
of the de Sitter radius, propagating as light wave fronts. Related
cosmological aspects are shortly discussed.
The analysis, performed for flat expanding universe, shows
exponential damping of the back-reaction effects far from these
spherical objects. The obtained solutions for the semiclassical
Einstein equations in de Sitter background
can be straightforwardly extended also to the anti-de Sitter geometry.
\end{abstract}

\vspace{7cm}

\centerline{{\footnotesize e-mails: FURSAEV@NA.INFN.IT and MIELE@NA.INFN.IT}}

\newpage
\baselineskip=.8cm

\section{Introduction}

In the modern picture of the early universe the exponential expansion
period is known to be successful in solving a number of important
problems occurring in the standard cosmology. During this inflationary
phase, described by de Sitter geometry, quantum field effects
played an essential role for subsequent formation of the present universe.
In this framework, the knowledge of the corresponding quantum state,
originating from Planck epoch, is important to reconstruct the
phase transitions pattern of the inflation. One of the possibilities generally
used \cite{a1}, \cite{a2} is to consider this quantum state as a de Sitter
invariant vacuum experienced by a freely moving observer as a thermal
bath at the Hawking temperature $\beta_{H}^{-1}$ \cite{a5}.

A wider class of quantum states, associated in static de Sitter
coordinates to a thermal equilibrium at arbitrary temperature $\beta^{-1}$,
has been investigated in our previous work \cite{c2}. In the present letter we
extend this analysis studying the back-reaction on de Sitter
geometry in these quantum states.

For simplicity we restrict our consideration to massless conformally invariant
scalar theory where the finite temperature Green function,
globally defined on the whole space-time, can
be exactly found. However, the corresponding
renormalized stress tensor for arbitrary temperatures
includes a traceless part singular on the horizon surface of two
antipodal static coordinate systems, where thermal equilibrium is introduced.
This extremely large energy density implies that, very close
to the horizon, physical quantities can be
computed only in the framework of quantum gravity, theory which would
provide a natural short distance cut-off \footnote{
Note that recently much attention has been paid to similar
divergences appearing in statistical mechanical computations
of black hole entropy \cite{c3}-\cite{c6}, when temperature different from the
Hawking one has to be introduced to obtain the derivative of the
partition function. Possible role of superstring
theory, providing such fundamental cut-off, is discussed in \cite{c5}.}.
Thus, the singular part of the stress tensor
can be interpreted as matter distribution sharply concentrated
over two spherical shells of radius $r_0$ (horizon size) that propagate
through the de Sitter hyperboloid as light wave fronts. So far as,
apart from the rest space-time,
the quantum field inside the domains bounded by the shells is in thermal
equilibrium, we will refer to these spherical 3-d
regions as {\it bubbles} \footnote{It is worth reminding that these  bubbles
are different from the topological defects nucleated during phase transitions},
that can be characterized by temperature $\beta^{-1}$, internal energy,
and entropy.

This letter is organized as follows: in section 2 we compute the renormalized
stress tensor at finite temperature $\beta \neq \beta_{H}$, and discuss
its properties inside and outside the bubbles.
In section 3 this tensor is used to
evaluate, in the linear approximation, the back-reaction effects.
The model of the flat expanding de Sitter universe is then
considered in section 4 to study the cosmological implications of the
chosen quantum state. Finally, we give our conclusions and remarks.

\section{The quantum state and stress tensor}

Let us determine the average value of the energy
momentum tensor for conformally invariant scalar field theory
$<\hat{T}_{\mu}^{\nu}>_{\beta}\equiv T_{\mu}^{\nu}(\beta)$ in static de Sitter
space, where $\beta$ is the inverse temperature of the system.
This tensor can be decomposed as
\begin{equation}
T_{\mu}^{\nu}(\beta)=\tilde{T}_{\mu}^{\nu}(\beta)+\frac 14 \delta
_{\mu}^{\nu}T~~~,
\label{eq:struc}
\end{equation}
where $\tilde{T}_{\mu}^{\nu}(\beta)$ is the traceless part, and
$T\equiv T_{\mu}^{\mu}(\beta)$ is
the local conformal anomaly that does not depend on the quantum
state of the system \cite{1}. In particular, for de Sitter space
\begin{equation}
T={1 \over 240\pi ^2 r_0^4}
\label{eq:rr}
\end{equation}
($r_{0}$ stands for the de Sitter radius), and due to the conservation law
$\tilde{T}_{\mu;\nu}^{\nu}=0$, the traceless part has
only one independent component. Consequently, to find all its components
it is sufficient to calculate only one of them.

It is worth reminding that static de Sitter metric $g_{\mu \nu}$ reads
\begin{equation}
ds^2= g_{\mu \nu} dx^{\mu}dx^{\nu}
=(1-r^2/r_0^2)dt^2-(1-r^2/r_0^2)^{-1}dr^2-r^2(d\theta ^2+\sin ^2\theta
d\varphi ^2)~~~,
\label{eq:metr}
\end{equation}
where $x^{\mu}\equiv(t,r,\theta,\varphi)$. In this space the temporal
component of the stress tensor $\tilde{T}_{t}^{t}(\beta)$
can be obtained observing that \cite{a8}
\begin{equation}
<\hat{T}^t_t>_{\beta}-<\hat{T}^t_t>_{\beta=\infty}=
{\pi^2 \over 30 }~ {\beta^{-4} \over g_{tt}^2}~~~.
\label{eq:subtr}
\end{equation}
The last equation can be also rewritten in the following form
\begin{equation}
<\hat{T}^t_t>_{\beta}=<\hat{T}^t_t>_{\beta _H} +
{\pi^2 \over 30 } \left( 1- {\beta^4 \over \beta_H^4} \right)
{\beta^{-4} \over g_{tt}^2}~~~,
\label{eq:subtr2}
\end{equation}
where $\beta_H \equiv 2\pi r_0$ denotes the inverse Hawking temperature.
Hence, due to the fact that for $\beta=\beta_{H}$
the stress
tensor is completely anomalous ($\tilde{T}_{\mu}^{\nu}(\beta_{H})=0$),
we have
\begin{equation}
\tilde{T}^t_t(\beta) = <\hat{T}^t_t>_{\beta}-<\hat{T}^t_t>_{\beta =\beta _H}=
{\pi^2 \over 30} \left( 1- {\beta^4 \over \beta_H^4}\right)
{\beta^{-4} \over g_{tt}^2}~~~.
\label{eq:density}
\end{equation}
This enables to obtain the stress tensor in the form
\begin{equation}
<\hat{T}^{\nu}_{\mu}>_{\beta}=
{\pi^2 \over 30} \left( 1- {\beta^4 \over \beta_H^4}\right)
{\beta^{-4} \over g_{tt}^2} diag\left(1,-\frac 13,-\frac 13,-\frac 13\right)+
{1 \over 960\pi ^2 r_0^4}\delta ^{\nu}_{\mu}~~~,
\label{eq:stresstensor}
\end{equation}
that coincides at zero temperature with the known result for the static
conformal vacuum \cite{2}.

As one can see from (\ref{eq:stresstensor}) the traceless part dominates
near the horizon and corresponds for $\beta~<~\beta_H$
to the energy momentum tensor of a gas
of massless scalar particles \cite{3}. In this case,
the energy density has the usual Planck form and results to be
\begin{equation}
\tilde{T}^t_t(\beta)={\pi ^2 \over 30} \tilde{\beta}^{-4}~~~,
\label{eq:energy}
\end{equation}
where $\tilde{\beta}^{-1}=\beta^{-1}( 1-\beta^4/\beta_{H}^4)^{1/4}
g_{tt}^{-1/2}$ plays the role of a local, redshifted temperature.
The similar structure for the energy density in finite temperature Rindler
space is given in \cite{4}.

Let us point out that (\ref{eq:stresstensor}) can be obtained
by the standard procedure from the finite temperature Green function
given by \cite{a8}
\begin{equation}
G_{\beta}(x,x') = i \frac{[(1 - r^ 2/r_0^2 )(1- r'^2/r_{0}^2)]^{-1/2}}
{2 \beta_{H} \beta \sinh\alpha_{1}}
\frac{\sinh(\alpha_{1} \beta_{H}/\beta)}
{\cosh(\alpha_{1} \beta_{H}/\beta)- \cosh[( t-t') \beta_{H}/r_{0} \beta]}~~~,
\label{eq:green}
\end{equation}
($\cosh\alpha_{1}=[(r_{0}^2 -r^2)(r_{0}^2 -r'^2)]^{-1/2}
\{ r_{0}^2 - r r'[\cos\theta \cos\theta' + \sin\theta \sin\theta'
\cos(\varphi -\varphi')]\}$),  with the subtraction of
its value corresponding to the
Hawking temperature. Although expression (\ref{eq:green})
is defined only in the region bounded by the horizon surface,
it can be extended to the total de Sitter space-time.
In terms of conformal
diagram, figure 1,
 where static coordinate system (\ref{eq:metr}) maps into the
region (a),  it is easy to prove that $G_{\beta}(x,x')$
can be smoothly continued from the region (a) into the casually connected
domains (c) or (d), by moving one by one its arguments through the horizon
surface. Then extending this procedure from (c) or (d) to (b) we obtain a
globally defined two-point Green function, smooth everywhere for not coinciding
arguments. Such procedure completely fixes in all the de
Sitter space-time the quantum state,
whose properties can be derived from the structure of
the associated global energy momentum tensor. This global stress tensor as
opposite to the Green function is singular on the horizon surface, which
means that close to the horizons our semiclassical approach
is not applicable and the complete theory of quantum
gravity should be used for computing the quantum effects.

As we have already mentioned one can interpret the energy singularity
as a matter distribution over the surface of two spherical bubbles
having the horizon size $r_0$, and propagating as fronts of light waves.
Regions (a) and (b) represent their internal regions,
where the stress tensor has identical forms,
(\ref{eq:stresstensor}).
In the external regions (c), (d) the metric (\ref{eq:metr}) changes to
\begin{equation}
ds^2=(r^2/r_0^2-1)^{-1}dr^2-(r^2/r_0^2-1)dt^2-r^2(d\theta ^2+\sin^2\theta d
\varphi ^2)~~~,
\label{eq:metr2}
\end{equation}
and describes an expanding (or shrinking) universe where $r$ now plays
the role of time
and $t$ has the meaning of a spatial coordinate. In these case, the
horizon surfaces are removed to spatial infinities ($t=\pm\infty$),
and the structure of the stress tensor can be found on the base of
(\ref{eq:stresstensor}). It follows immediately that in these domains
$\tilde{T}_{\mu}^{\nu}(\beta)$ corresponds to a non equilibrium thermal
system. Although the energy flows from the infinities
($t=\pm\infty$) are absent, the energy density decreases to the vacuum one
due
to the factor $[g_{tt}(r)]^{-2}$ in according with adiabatic expansion
of this universe.

A remark concerning the thermodynamical parameters of the bubbles
like their energy, and entropy it is worth being done here.
These quantities can be computed in two ways: through the local energy
$T^t_t(\beta)$ \cite{c4}, or in terms of the one-loop partition function
\cite{c2}. The first method cannot be straightforwardly applied
due to the divergence of the local energy and entropy at the horizon,
whereas the other one gives a finite answer under $\zeta$-function
regularization \cite{c2}. This fact might be understood in terms
of renormalization of bubbles surface energy, which can cancel the
local divergence. To partially support the above statement we refer to the
analysis carried out for manifolds with conical singularities \cite{a19}.
Such kind of defects is inherent to "gravitational instantons" associated
to static de Sitter space (\ref{eq:metr}) at $\beta \neq \beta_{H}$ \cite{c2}.

\section{The semiclassical analysis}

Consider now changing of de Sitter geometry induced
by the vacuum polarization in the given quantum states. Einstein
equations in semiclassical approximation are known to read as \cite{1}
\begin{equation}
R_{\mu \nu}(\bar{g}) - \frac{1}{2} R(\bar{g}) \bar{g}_{\mu \nu}
+ \Lambda \bar{g}_{\mu \nu} + \alpha ~{}^{(1)}H_{\mu \nu}(\bar{g})~+
\beta~{}^{(2)}H_{\mu \nu}(\bar{g})
 = - 8 \pi G <\hat{T}_{\mu \nu}>_{\beta}~~~.
\label{eq:einstot}
\end{equation}
Here $\bar{g}_{\mu\nu}=g_{\mu\nu}+h_{\mu\nu}$ indicates the solution of
(\ref{eq:einstot}) including quantum corrections $h_{\mu\nu}$ to de Sitter
metric $g_{\mu\nu}$, and $\Lambda =3/r_0^2$ is the cosmological constant.
As for the additional
tensors ${}^{(1)}H_{\mu \nu}$ and ${}^{(2)}H_{\mu \nu}$ that appear due
to the quantum corrections, and are defined
in terms of the geometrical quantities \cite{1}, we will neglect them in the
further analysis assuming the unknown constants $\alpha$, $\beta$ in
(\ref{eq:einstot}) to be equal to zero.
Then, expanding (\ref{eq:einstot}) up to the first order in the metric
perturbation $h_{\mu\nu}$, we have
\begin{eqnarray}
 h^{\alpha}_{\mu;\nu\alpha}+h^{\alpha}_{\nu;\mu\alpha}-
h^{\alpha}_{\alpha;\mu\nu}  - h^{~~;\alpha}_{\mu\nu;\alpha}
+ {6 \over r_0^2} h_{\mu \nu} +
\left(h^{\beta;\alpha}_{\beta;\alpha} -
h^{\beta;\alpha}_{\alpha;\beta}- { 3 \over r_0^2} h^{\alpha}_{\alpha}\right)
g_{\mu \nu} =16 \pi G <\hat{T}_{\mu \nu}>_{\beta}~.
\nonumber\\
\label{eq:eins1}
\end{eqnarray}

An essential progress in solving these equations can be achieved observing
that the anomalous part of the renormalized stress tensor in the r.h.s. of
(\ref{eq:eins1}) only gives rise to a redefinition of the de Sitter
radius $r_{0}$. Taking into account this trivial effect we can
use instead of
the total tensor only its traceless part $\tilde{T}_{\mu}^{\nu}(\beta)$.
Thus, after a little transformation, the initial equations take the form
\begin{eqnarray}
h_{\mu\nu;\alpha}^{~~;\alpha} + {2 \over r_0^2} h_{\mu \nu}
= - 16 \pi G~ \tilde{T}_{\mu \nu}({\beta})~~~,
\label{eq:eins5}
\end{eqnarray}
where we imposed the gauge conditions $h_{\mu\nu}^{~~;\nu}=0$, $h_{\mu}^
{\mu}=0$ (the latter is compatible with the dynamical equations and fixes
the residual gauge freedom). Besides this, the property
of de Sitter space for which $R_{\lambda\nu\mu\alpha}=
r_0^{-2}(g_{\lambda\mu}g_{\nu\alpha}-g_{\lambda\alpha}g_{\nu\mu})$ has been
used in (\ref{eq:eins5}).

To proceed in our analysis, it is worth reminding that in the
given quantum state ($\beta \neq \beta_{H}$)
the total de Sitter symmetry $SO(1,4)$ is broken to the
subgroup $T_1\times SO(3)$, where $T_1$ stands for de Sitter boosts associated
with translations along the coordinate $t$.
This enables us to
restrict the degrees of freedom of $h_{\mu \nu}$, to the only nonzero
components
$h_{tt}(r)$, $h_{rr}(r)$, $h_{tr}(r)=h_{rt}(r)$, and $h_{ik}\equiv
-a(r)r^2\gamma _{ik}$,
where  the indexes $i,k$ are referred to the $\theta,\varphi$  coordinates,
and $\gamma _{ik}$
is the metric on $S^2$ (in accordance with
(\ref{eq:metr}) $g_{ik}=-r^2\gamma _{ik}$).

Due to the residual symmetry and
gauge conditions imposed, there is only one independent equation among
(\ref{eq:eins5}). After some algebra one can show that in static coordinates
it reduces to a second order differential equation for $h_{rr}$
\begin{equation}
\left[ 4\left( 1 - y \right) \frac{d^2}{dy^2}
- {26 y -16 \over y} \frac{d}{dy} -
{28y-8 \over y^2} \right]h_{rr}(y) = - \frac{K(\beta)}{y^4} ~~~,
\label{eq:redfeq}
\end{equation}
written in terms of the variable $y(r)\equiv1-r^2/r_0^2$ and a constant
\begin{equation}
K(\beta) \equiv {G \over 90 \pi r_{0}^2}\left(1-{\beta_H^4 \over \beta ^4}
\right)~~~.
\label{eq:defk}
\end{equation}
Remarkably, the homogeneous
equation associated to (\ref{eq:redfeq}) admits two simple solutions:
$y^{-2}$,  and $[y^2(1-y)^{3/2}]^{-1}$. Thus
the general integral of (\ref{eq:redfeq}) can be given in the
following form
\begin{eqnarray}
h_{rr}(r) &= &{K(\beta) \over 4y^2(r)}\left\{ { 2 \over 3}+
{2 \over 1-y(r)} +{1 \over [1-y(r)]^{3/2}}\log
\left[{1-\sqrt{1-y(r)} \over 1+\sqrt{1-y(r)}}\right]\right\}
\nonumber\\
&+& {A \over y^2(r)} +{ B \over y^2(r)[1-y(r)]^{3/2}}~~~.
\label{eq:int}
\end{eqnarray}
It is worth mentioning that coordinate systems in de Sitter spaces,
(\ref{eq:metr}), (\ref{eq:metr2}),
can be treated on equal footing if one adds a small imaginary part to
de Sitter radius $r_0\rightarrow r_0+i\epsilon$. This
regularization preserves the structure of all the equations
in both quantum and classical theory. It removes the singularity at the
horizons in such a way that (\ref{eq:metr2}), and (\ref{eq:metr})
can be unified in a single expression for the
metric valid for $0\le r < \infty$. However,
although such regularization allows to define the solution in all the region
$0\le r <\infty$, the integration constants $A$ and $B$
cannot be chosen to be equal in all this interval.
Indeed, if $A$ and $B$ are fixed in some way in the inner domain (a), then
after
passing in the external region, say (c), one obtains due to the logarithm
complex values for the function
$h_{rr}(r)$. It means that integration constants for external and internal
problems in the given semiclassical approach should be found independently,
on the base of additional physical motivations.

In the static regions $A$ and $B$ can be chosen from the
condition $h_{\mu\nu}(r)=0$ at $r=0$ that corresponds to the natural
assumption for the vacuum effects to disappear in the flat limit
$r/ r_0 << 1$, when the background space converts into Minkowski one.
As for the external domains, one can require for the perturbed metric to
approach the de Sitter metric in the limit $r\rightarrow \infty$, where
local quantum effects disappear (see (\ref{eq:stresstensor})). The
components $h_{tt}$, $h_{ik}$, can be obtained from (\ref{eq:int}) using
constraints. On the other hand, the non diagonal element $h_{tr}$,
fixed by the condition $h_{t\nu}^{~~;\nu}=0$, turns out to
be completely independent of the stress tensor and other components. For this
reason, we put it to be zero everywhere, which inside the horizon can be also
justified by claiming its regularity at $r=0$.

By virtue of the expression (\ref{eq:int}) and conditions chosen
we can represent all the non zero components of the metric perturbations
in the following form
\begin{eqnarray}
h_{tt}(r) & = & {K(\beta) \over 4} \left[ 2 -{4r^2 \over 3r_{0}^2} +
{ r_{0} \over 2r}
\log\left({r_{0} - r \over r_{0} + r }\right)^2\right]+A\left(3-
{2r^2 \over r_0^2}\right)+B{r \over r_0}~~~,
\label{eq:solfin1}\\
\nonumber\\
h_{rr}(r) & = &
{ K(\beta) \over 4~y^2(r)} \left[\frac 23 + {2r_{0}^2 \over r^2}+
{r_0^3 \over 2r^3}
\log\left({r_{0} - r \over r_{0} + r}\right)^2 \right]+{A \over y^2(r)}
+B {r_0^3 \over y^2(r)r^3}~~~,
\label{eq:solfin3}\\
\nonumber\\
a(r) & = & { K(\beta) \over 4 }\left[ -\frac 23 + {r_0^2 \over r^2 ~y(r)}
+ { r_{0}^3 \over 4 r^3} \log\left({r_{0} - r \over r_{0} + r }\right)^2\right]
-A+B {r_0^3 \over 2r^3}
{}~~~.
\label{eq:solfin4}
\end{eqnarray}

Following the previously mentioned requirements,
the integration constants  $A$ and $B$ have to be put equal to zero
for $0\le r<r_0$, whereas for $r_0<r<\infty$ we have $A=-K(\beta)/6$.
As far as the value of $B$ in the external region is concerned,
it is left undetermined.
This fact can be understood reminding that in the approximated
approach of semiclassical theory
two regions are separated by an infinite barrier,
whereas in the correct theory both solutions should be matched at
the horizon, so fixing the remaining constant. Note also that
solutions (\ref{eq:solfin1})-(\ref{eq:solfin4}) are valid only in the
region where the perturbation expansion is reliable. In terms of
the Planck length $l_{Pl} \equiv \sqrt{G}$, this region results to be
$|r -r_{0}| >> | 1 - \beta_{H}^4/\beta^4|^{1/2} l_{Pl}$. Therefore, if the de
Sitter radius $r_{0} \approx l_{Pl}$ the present analysis is not applicable
to the bubble interior.

Finally, we observe that solutions (\ref{eq:solfin1})-(\ref{eq:solfin4})
of semiclassical Einstein equations in de Sitter background, can be
immediately extended to anti-de Sitter space just substituting $r_{0}$
with $i r_{0}$. This is connected to the fact that the renormalized
stress tensor at finite temperature in anti-de Sitter space has the same
structure of de Sitter one, once a complex scalar field with Neumann and
Dirichlet boundary conditions for its two independent
components is chosen \cite{a20}.

\section{Gravitational effects }

We discuss now the properties of the space taking into account
quantum corrections (\ref{eq:solfin1}) - (\ref{eq:solfin4}) to de Sitter
metric. It can be done considering the different models of de Sitter
space corresponding to flat, open or closed expanding universes.
For sake of simplicity we restrict in this letter our analysis to the flat
model, which is generally used for cosmological applications.
A more complete study will be given in a forthcoming paper \cite{5}.

The corresponding metric has the familiar form
\begin{equation}
ds^2=d\tau ^2-e^{2\tau /r_0}\left(d\xi^2+\xi^2d\Omega ^2\right)~~~,~~~0
\leq\xi<\infty~~~,
\label{eq:flatun}
\end{equation}
($d\Omega ^2=d\theta ^2+\sin ^2\theta d\varphi ^2$ is the line element
on a sphere). This system of coordinates maps only a half of the de
Sitter hyperboloid. For a more clear representation we assume it
to cover regions (a) and (c) on figure 1. In this case, one of
the singularities of the global stress tensor is placed inside
the universe (\ref{eq:flatun}), in such a way that bubble's center
corresponds to the origin of coordinates. At the same time,
the other one is at spatial infinity ($\xi \rightarrow \infty$).
The map between (\ref{eq:flatun}) and (\ref{eq:metr}) or (\ref{eq:metr2})
can be written as follows
\begin{equation}
t=\tau - {r_0 \over 4}\log\left(1-{\xi ^2 \over r_0 ^2}e^{2\tau /r_0}\right)^2
{}~~~,~~~r=\xi e^{\tau /r_0}~~~.
\label{eq:map}
\end{equation}
Note that by definition, the shell of the bubble coincides with de Sitter
horizon thus its size remains unchanged during inflation
\footnote{A discussion about
this point is given in section 5.}, but due to the expansion every point,
once inside it, will move away. Near the matter distribution surface
$\xi_{0} = r_{0} ~\exp(-\tau/r_{0})$ the metric (\ref{eq:flatun})
changes to
\begin{equation}
ds^2=(1+h_{\eta \eta})~d\eta^2+ 2 h_{\eta\xi}~d\eta d\xi-\left(e^{2\eta/r_0}-
h_{\xi\xi}\right)d\xi ^2
- e^{2\eta/r_0}\xi^2\left(1+a \right) d\Omega ^2~~~,
\label{eq:flatun'}
\end{equation}
where we introduced $\eta$ instead of $\tau$ to indicate that now it is not
the proper time.
The metric perturbations $h_{\eta \eta}$, $h_{\eta \xi}$ and $h_{\xi\xi}$
follow from (\ref{eq:solfin1}), (\ref{eq:solfin3}) and map (\ref{eq:map}),
where $\tau$ is replaced by $\eta$
\begin{eqnarray}
h_{\eta \eta}(\eta,\xi) & = & {1 \over y^2}
h_{tt} +{\xi ^2 \over r_0^2}e^{2 \eta/r_0}h_{rr}~~~,
\label{eq:sol00}\\
\nonumber\\
h_{\eta\xi}(\eta,\xi) & = & {\xi \over r_0}e^{2 \eta/r_0}
\left({1 \over y^2} h_{tt} + h_{rr}\right)~~~,
\label{eq:sol0x}\\
\nonumber\\
h_{\xi\xi}(\eta,\xi) & = & e^{2\eta/r_0}\left({\xi ^2 \over r_0^2~ y^2}
e^{2 \eta/r_0} h_{tt} + h_{rr}\right)~~~.
\label{eq:solxx}
\end{eqnarray}
In Eqs. (\ref{eq:flatun'})-(\ref{eq:solxx}) all the functions $y$, $a$,
$h_{tt}$
and $h_{rr}$ are supposed to be expressed in terms of $\eta$ and $\xi$.
So far as $\eta$ does not represent the proper time we need an additional
transformation to pass to comoving coordinates  ($\tau,x,\theta,\varphi$),
similar to (\ref{eq:flatun}),
\begin{equation}
ds^2=d\tau ^2-e^{2\tau /r_0}\left[(1+ h_{xx}(x,\tau))dx^2
+x^2(1+h_{\Omega \Omega}(x,\tau))d\Omega ^2\right]\equiv d\tau^2 -
\gamma_{ij}dx^{i} dx^{j}
\label{eq:flatun''}
\end{equation}
with $i,j=1,2,3$, corresponding to $x$, $\theta$ and $\varphi$
respectively. The above transformation reads
\begin{eqnarray}
\eta  =  \tau + \eta_{1}(\tau,x)~~~~~,~~~~~
\xi  =  x + \xi_{1}(\tau,x)~~~,
\label{eq:deviations}
\end{eqnarray}
and the functions $\eta_{1}$ and $\xi_{1}$ obey to the following differential
equations
\begin{equation}
2{\partial \eta_{1} \over \partial \tau} + h_{\eta \eta}=0~~~~~,~~~~~
e^{2\tau/r_0}{\partial \xi_{1} \over \partial \tau}
-h_{\eta\xi}-{\partial \eta_{1} \over \partial x}=0~~~.
\label{eq:comov}
\end{equation}
In terms of these functions the metric corrections $h_{xx}$
and $h_{\Omega \Omega}$ result to be
\begin{eqnarray}
h_{xx} = 2{\eta_{1} \over r_0}+2{\partial \xi_{1} \over \partial x} -
e^{-2\tau/r_0} h_{\xi\xi}~~~~~,~~~~~
h_{\Omega \Omega}  = 2{\eta_{1} \over r_0}+2{\xi_{1} \over x} + a~~~.
\label{eq:urho}
\end{eqnarray}
It is worth observing that the integration constant for $\xi_{1}$, obtained
from (\ref{eq:comov}), is not relevant because it can always be
changed by redefinition of $x$ coordinate in (\ref{eq:flatun''}). As far as
$\eta_{1}$ is concerned, its value is fixed, up to an inessential numerical
constant, by requiring that metric
components (\ref{eq:flatun''}) depend only on the distance $ r = x\exp(
\tau/r_{0})$ from the center of the perturbation.

The metric components can be readily obtained from (\ref{eq:comov}) on the
total
space ($0 \leq r <r_{0}$ and $r_{0} < r < \infty$). For sake of brevity we
only present the expressions for the outside region to investigate the
metric asymptotic at infinity. They read
\begin{equation}
h_{\Omega\Omega}=K(\beta)\left[-\frac 54\log\left({r^2 \over r^2-r_0^2}\right)
+{r_0^3 \over 8r^3}\log\left({r-r_0 \over r+r_0}\right)+
{r_0^2 \over 4r^2}{2r_0^2-3r^2 \over r_0^2-r^2}\right]
{}~~~,
\label{eq:omom}
\end{equation}
\begin{equation}
h_{xx}=K(\beta)\left[-\frac 54\log\left({r^2 \over r^2-r_0^2}\right)+
{r_0^2 \over 4r^2}{r_0^2-3r^2 \over r_0^2-r^2}\right]~~~,
\label{eq:xx}
\end{equation}
where the undefined constant $B$ was neglected, because it does not affect
the long-range behavior ($r \gg r_{0}$) of perturbations. By
using (\ref{eq:omom}) and (\ref{eq:xx})
one can see that gravitational field far from the bubbles
decreases like $r^{-2}$. It corresponds, due to the expansion,
to an exponential falling with time $\tau$ for a point of given $x$.

The effects of the quantum corrections to the
metric (\ref{eq:flatun}) can be expressed in another way by computing the
spatial curvature $R^{(3)}(\gamma)$ of the universe (\ref{eq:flatun''})
\begin{equation}
R^{(3)}(\gamma)=h^{i;j}_{i;j}-h^{j;i}_{i;j}~^i={2 \over x^2} e^{-2\tau/r_0}
{d \over dx}(xf(x,\tau))~~~,
\label{eq:scurv}
\end{equation}
\begin{equation}
f(x,\tau)\equiv{d \over dx}(x~h_{\Omega\Omega})-h_{xx}=3a+x{d \over dx}a~~~.
\label{eq:f}
\end{equation}
Remarkably,
due to (\ref{eq:f}) it turns out to be completely independent of the constant
$B$ and takes the simple analytical forms
\begin{equation}
R^{(3)}(\gamma)|_{r<r_0}=K(\beta){3r_0^2-r^2 \over (r^2-r_0^2)^2}~~~,~~~
R^{(3)}(\gamma)|_{r>r_0}=K(\beta){r_0^2 \over r^2}
{r^2+r_0^2 \over (r^2-r_0^2)^2}~~~,
\label{eq:scurv2}
\end{equation}
for inside and outside regions respectively. By
definition (\ref{eq:defk})
of $K(\beta)$ it follows that curvature is negative
if the bubble temperature $\beta ^{-1}$ higher then the Hawking one
$\beta _H^{-1}$,
and changes the sign when $\beta ^{-1}<\beta _H^{-1}$. Moreover, its value at
$r=0$ does not depend on time $R^{(3)}(\gamma)|_{r=0} =
3 K(\beta) r_{0}^{-2}$, whereas
in external region $R^{(3)}$ decreases as $r^{-4}$.
This long range behaviour of the gravitational field is connected to
the massless nature of the chosen scalar field. For massive matter fields
one can expect that the gravitational effects will be exponentially
smeared, on a length of the order of the inverse mass,
outside the bubble shell.

\section{Conclusions and remarks}

In this letter the local properties of thermal states for a quantum
field in de Sitter space-time have been investigated. We interpret
the known singularity of renormalized stress-tensor as a matter
distribution located on two spherical surfaces (boundaries of bubbles)
moving far away one from each other. Such domains, characterized by
temperature, entropy, and etc., give rise to a number of semiclassical
effects.

To find the gravitational field produced by these perturbations,
the linearized Einstein equations on de Sitter background
have been exactly solved for stress tensor of a
conformal scalar field at finite temperature. The computations show
the gravitational effects are weak far from the bubbles, and
in the expanding universe they are exponentially damped.

In realistic cosmological models both the radius and thermodynamical
properties of the bubbles might change during the
universe evolution. Note that external region of static coordinates
(\ref{eq:metr}) can be considered as a "black hole" absorbing "information"
from the inner domain due to decays there of unstable field configurations
of the Higgs scalars.
As a result of this process, the bubble size,
associated to the horizon, grows up and achieves cosmological values
determined by the inverse of the Hubble constant, $H^{-1} \approx 10^{28}~cm$
at the present. However, due to the exponential damping, the presence of
these inhomogeneous regions does not seem to affect the observable part of
the universe, provided if it is far enough  from them.
We confine the present letter to these short remarks, leaving more
complete analysis of the cosmological aspects of such theory to
a forthcoming paper \cite{5}.

Finally, it is worth observing that
the solutions obtained for the semiclassical Einstein equations in de Sitter
background can be straightforwardly extended to the anti-de Sitter geometry.
In this case in fact, the stress tensor has the same structure of
the de Sitter one, once a complex scalar field with appropriate boundary
conditions is chosen \cite{a20}.

We would like to thank Bruce Allen and Sergey Solodukhin for their
interesting remarks. This work was supported in part by International Science
Foundation (Soros) Grant No. Ph1-0802-0920.

\newpage

{\bf Figure caption}

Figure 1: In this diagram left and right edges have to be identified.
Regions (a) and (b) correspond to static coordinates with $0 \leq r
< r_{0}$, separated by light cones from (c) and (d) regions where
$r_{0} < r < +\infty$. Dashed lines in regions (a) and (b) represent the
trajectories of a particle and its antipod being in the coordinate origins.
\end{document}